\documentclass[twoside,fleqn]{article}

\usepackage{espcrc2,psfig}
\newcommand{\AmS}{{\protect\the\textfont2
  A\kern-.1667em\lower.5ex\hbox{M}\kern-.125emS}}
\newcommand{\gev}{\mbox{GeV}}
\newcommand{\mev}{\mbox{MeV}}
\newcommand{\fig}{\mbox{Fig.}}
\newcommand{\imag}{\mbox{Im}\,}
\newcommand{\real}{\mbox{Re}\,}
\newcommand{\degrees}{^o}
\newcommand{\pepe}{\rm P.P.}
\newcommand{\dd}{\rm d}

\title{Chiral-dispersive calculations
 of $\pi\pi$ scattering confront experiment\footnotemark[1]}    
\author{
J.R.Pel\'aez
\address{
Dept. de F\'{\i}sica Te\'orica II, Univ. Complutense de Madrid.
28040 Madrid. Spain} and F. J. 
Yndur\'ain\address{
Dept. de F\'{\i}sica Te\'orica, C-XI
 Univ. Aut\'onoma de Madrid,
 Canto Blanco,
E-28049, Madrid, Spain.}
}

\begin{document}

\begin{abstract}
In a series of papers  
we have applied 
several  sum rules and  forward dispersion relations, 
to $\pi\pi$ scattering.
We have found that some widely used data sets fail to
satisfy these constraints, and
we have provided an amplitude
that describes data consistently with the dispersive tests.
Furthermore, we noted that the input and precision claimed 
in a Roy equation analysis by
Colangelo, Gasser and Leutwyler (CGL), 
lead to several mismatches with some sum rules. 
Subsequently,
Caprini, Colangelo, Gasser and Leutwyler claimed 
that our Regge parametrization was incorrect.
We collect here
the answers to their various claims, 
and try to clarify the points of agreement and disagreement,
showing {\sl experimental} 
evidence that substantiates our results,  and that, in addition, 
their representation fails to satisfy all three forward dispersion relations 
up to $\sqrt{s}\leq800\, \mev$ by several standard deviations.
\end{abstract}

\maketitle

\pagestyle{empty}
\arraycolsep=0.5pt
\section*{Introduction}

\renewcommand{\thefootnote}{\fnsymbol{footnote}}
\footnotetext[1]{Joint contribution of the two talks
presented by the authors in  QCD05, Montpellier, France, July 2005.}
Some time ago, Ananthanarayan, Colangelo, Gasser and Leutwyler \cite{1} (ACGL)
and Colangelo, Gasser and Leut\-wyler \cite{2} (CGL) have used 
experimental information, unitarity, analyticity
(in the form of  
the Roy equations) and, in CGL, chiral perturbation theory, 
to construct the  $\pi\pi$ 
scattering amplitude. In CGL,
an outstanding precision was claimed, at the percent level,
for scattering lengths and effective ranges
 of S, P, D and F waves. In addition, 
CGL provided parametrizations for the  S, P phase shifts 
up to $s^{1/2}=0.8\,\gev$. 

In a series of works, referred to as PY1,\cite{3}
PY-Regge,\cite{4} PY-Sardinia\cite{5} and  PY-FDR\cite{6} (collectively, PY) 
we have contested the  {\it input} of both ACGL and CGL for the following reasons:
\newcounter{input}
\begin{list}{\roman{input}.}{\usecounter{input}
\setlength{\leftmargin}{0.4cm}
\setlength{\labelsep}{0.cm}}

\vspace{-.1cm}
\item  The high energy ($s^{1/2}>1.4\,\gev$), in
    particular the Regge parametrization, is incompatible
with data 
and violates factorization.
\item A relevant source of uncertainty, the error on the phase
    $\delta_0^{(0)}$ at 800~\mev, a crucial input for the matching between
low and high energy representations, is
    largely underestimated. The central value of either $\delta_0^{(0)}$
or $\delta_0^{(2)}$ is inconsistent with forward dispersion relations.

\item The D2 wave disagrees,
    both at low and high energy, with what is found from experiment.
\end{list}

This shed doubts on the precision claimed in CGL (final uncertainties 
in ACGL are larger). In fact, using experimental input, we showed that
\begin{list}{\roman{input}.}{\usecounter{input}
\setlength{\leftmargin}{0.4cm}
\setlength{\labelsep}{0.cm}}
\item[iv.] Some low energy parameters given in CGL,
    mainly those for D waves, do not
satisfy the Froissart Gribov sum rules.  
The P wave effective range parameter, $b_1$,
deviates by several standard deviations from what one finds
    from the pion form factor \cite{7} or from a sum rule.
\item[v.] The CGL final amplitudes fail to verify forward
    dispersion relations up to 800~\mev, again  by several standard deviations.
\end{list}

In the time between PY1 and PY-Regge, Caprini, 
Colangelo, Gasser and Leutwyler\cite{8} (CCGL) 
claimed to counter
the criticism in PY1. 
These claims where answered in PY-Regge and PY-FDR 
(and PY-Sardinia and \cite{9}).
However, in this note we provide a more formal risposte 
collecting all arguments together and also
including comments on a recent article of Colangelo \cite{10}.

Let us first enumerate the main points in which CCGL claim to 
answer the criticism of PY1:
\begin{list}{\arabic{input}.}{\usecounter{input}
\setlength{\leftmargin}{0.3cm}
\setlength{\labelsep}{0.cm}}
\item The asymptotic behaviour used in PY1. CCGL still
    consider it inferior to that used in ACGL and CGL in that it is
    not in ``equilibrium" with low energy, and that our asymptotics ``violates crossing rather strongly''.
\item
Based only on a selection of data,
CCGL conclude that CGL describes the S0 data,
whereas our ``tentative solution''does not.
\item CCGL claim that, even with PY1 Regge formulas, the
    CGL results are essentially unchanged.
\item CCGL consider the {\sl Olsson sum
      rule} and agree that it is not satisfied by CGL, if using the
    Regge asymptotics of PY1.  From this, CCGL conclude that ``the
    asymptotics used in PY1 is inconsistent with the theoretical
 predictions for    S-wave scattering lengths''.
\item  They keep their  $b_1$ prediction claiming that
the Froissart Gribov sum rule is subject to large uncertainties
due to the Regge representation.
\end{list}

\section{Regge asymptotics
and the D2 wave}

\noindent
The ACGL and CGL Regge  violates
factorization, 
a well-known property of Regge theory, for all trajectories,
but, in particular,
for the Pomeron exchange 
by a factor larger than two.
 
The Regge parameters in ACGL, CGL are 
obtained by ``balancing'' the 
high and low energy contributions in a number of crossing 
sum rules. CCGL conclude that the Regge representation 
in PY1 is {\sl incompatible with crossing 
symmetry} because it does not satisfy  
crossing sum rules, that are better satisfied by 
their Regge representation ( which they 
concede that could be improved).

However, although ignored then by ACGL, CGL, CCGL and ourselves
(we remedied this in PY-Regge), there are many data on 
$\pi^0\pi^-$ and, particularly, on 
$\pi^-\pi^-$ and $\pi^+\pi^-$ total cross sections at high energy\cite{14}.
This is shown in \fig~1 where it is clearly seen that the 
Regge expressions 
used by ACGL, CGL and CCGL are systematically
 below the data whereas our parametrization
falls on top of them. 
In fact, in PY-Regge, we used the $\pi\pi$ experimental data, 
together with the very 
precise $\pi N$, $KN$ and $NN$ cross sections data to get 
an accurate verification of 
factorization and a precise Regge parametrization. In
the Appendix~B of PY-FDR, 
by means of the Froissart Gribov representation,
we refined the $\rho$ residue and slope obtaining a value compatible
with our previous determinations.

\begin{figure}%[htb]
%\vspace{.3cm}
\hbox{\psfig{file=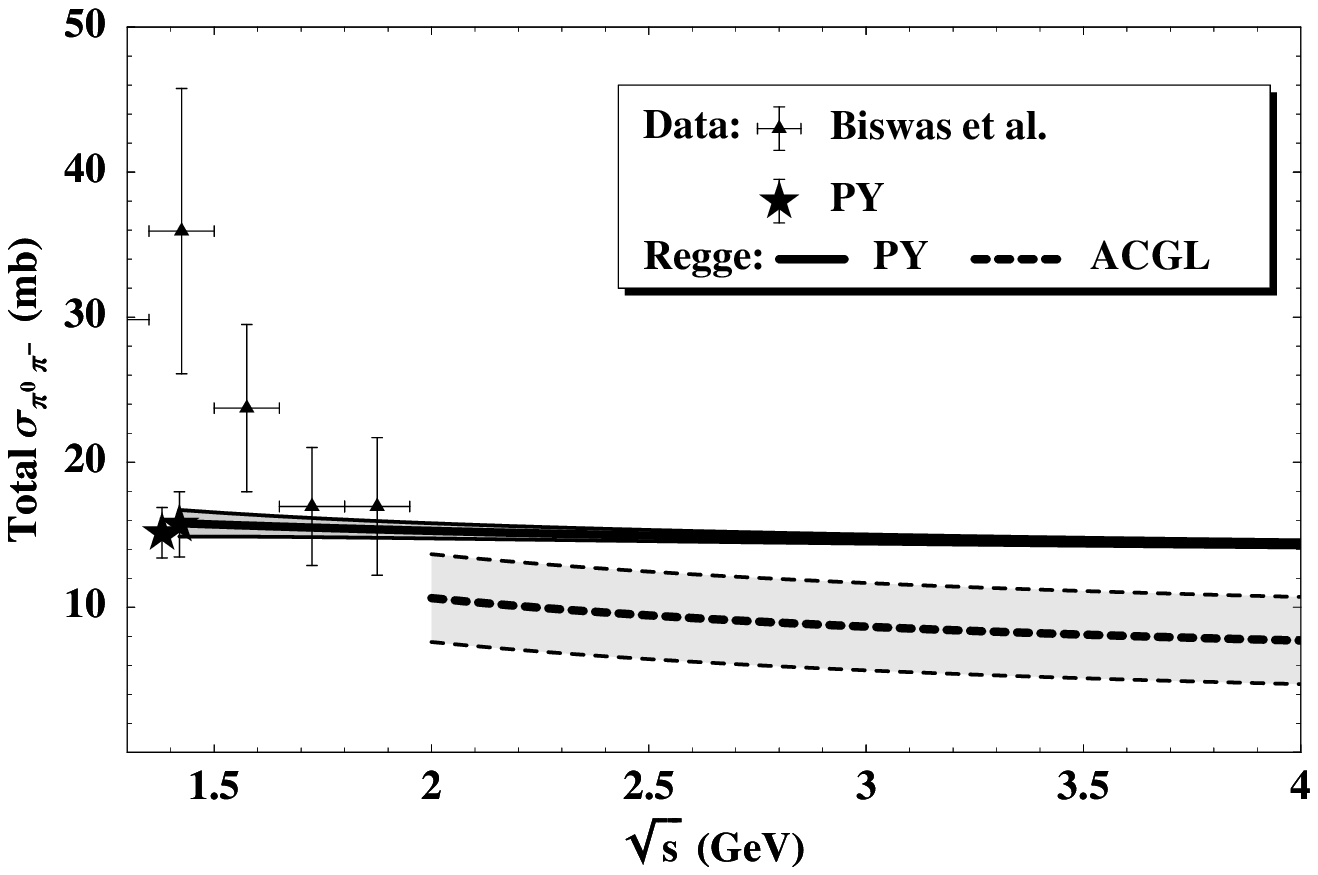,width=3.8cm}\hspace{-.2cm}
\psfig{file=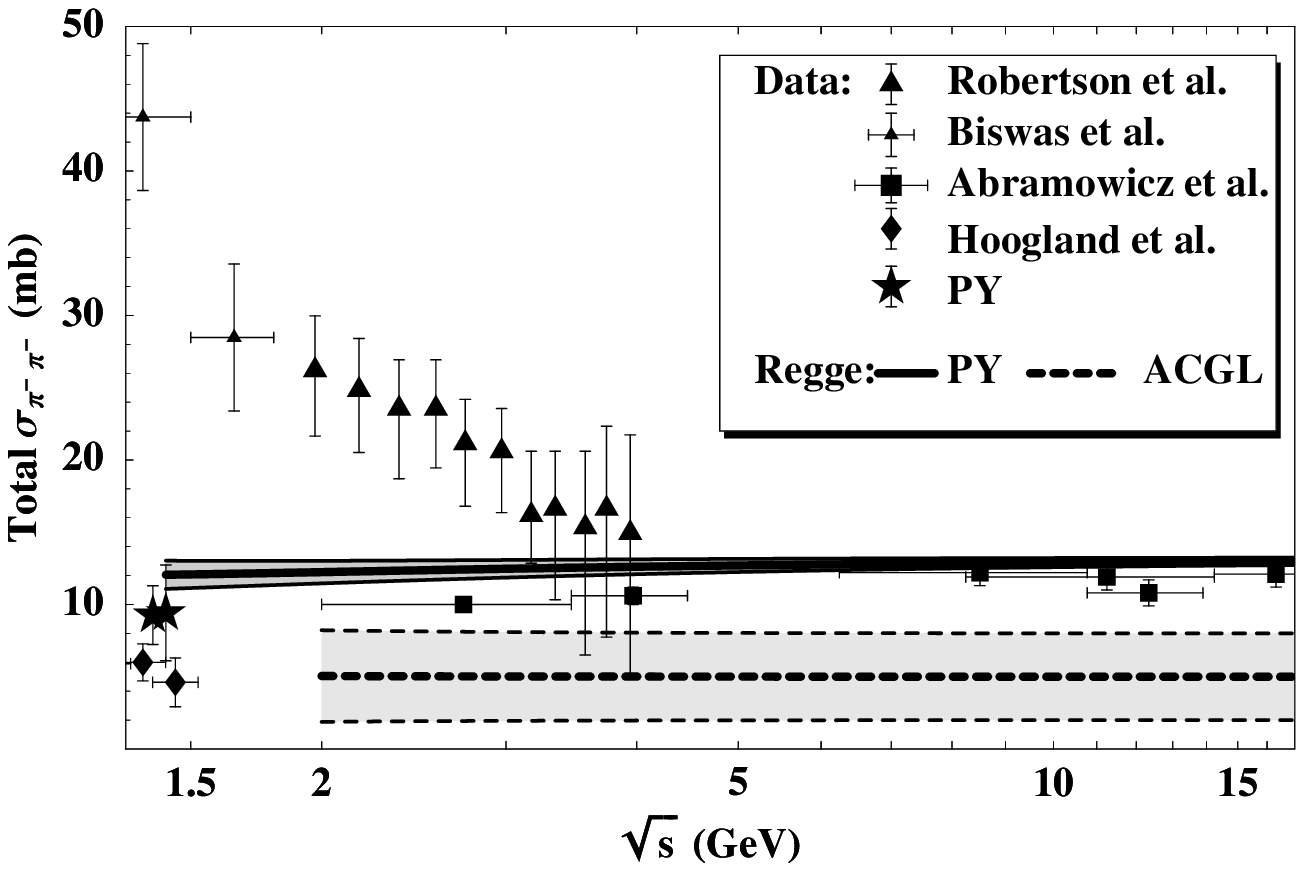,width=3.8cm}}
\psfig{file=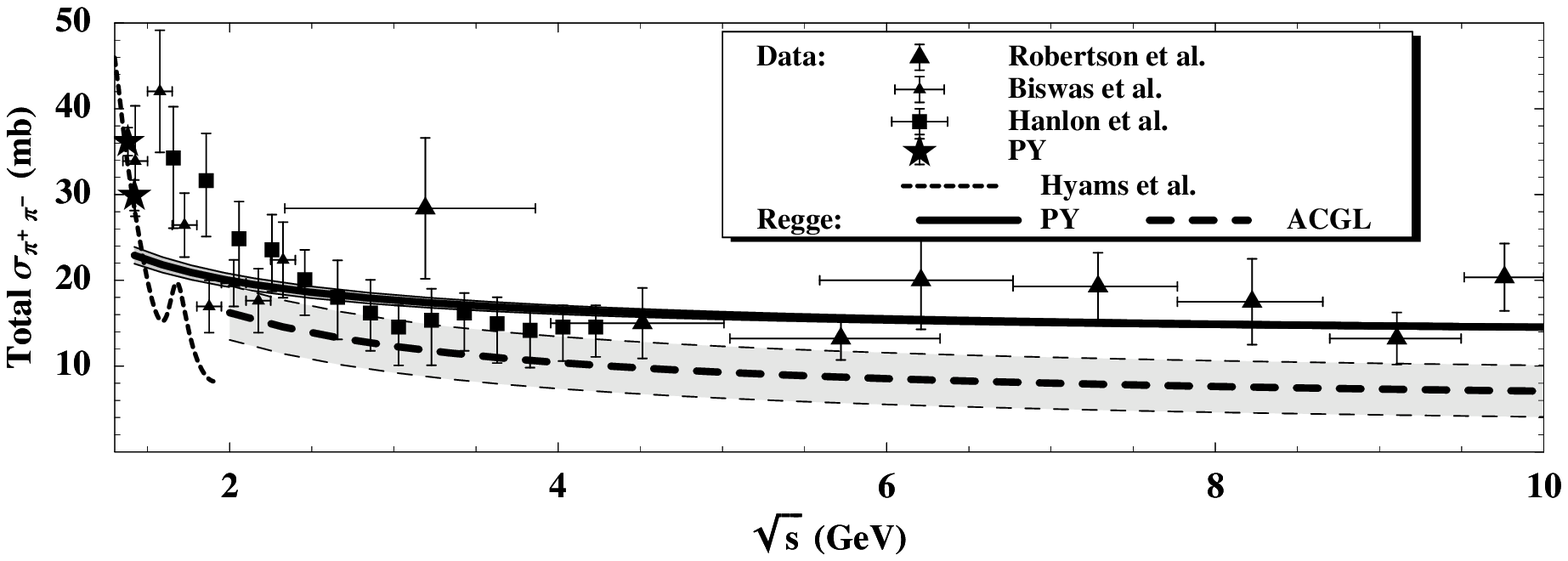,width=7.6cm}
{\footnotesize {\bf Figure 1} \rm $\pi\pi$ total cross section data \cite{14}. 
The stars (PY) are reconstructed in PY-Regge from experimental 
phase shift analysis. 
Continuous lines (PY): Regge parameters as in 
PY-Regge, in agreement, within errors with PY1. 
Dashed lines: ACGL Regge representation.
The grey bands cover the respective uncertainties.
The dotted line is the reconstructed $\pi^+\pi^-$ cross section from 
the Cern--Munich analysis; (\fig~7 in \cite{15a})
}
%\vspace{.3cm}
\end{figure}

Although the data supports our Regge parametrization,
one might still wonder about the sum rules. 
However \cite{9}
these sum rules are (A) fairly well
satisfied by the PY1 representation, 
{\sl when errors are taken into account}, which are notoriously absent in Eqs.~(11) 
and the next equation in CCGL; and, (B)  
At low energies the 
S-wave contribution
cancels, and, in some cases, also the P wave is absent. 
Thus, these sum rules are dominated by the D waves. 
But, for the D2 wave, ACGL and CGL borrow
the old fit from the book of
Martin, Morgan and Shaw \cite{12}
\begin{equation}
\delta_2^{(2)}(s)=-0.003(s/4M^2_\pi)\left(1-4M^2_\pi/s\right)^{5/2}.
\label{eq:(1.1)}
\end{equation}
Although it was
obtained only from data in the $0.625\gev\leq s^{1/2}\leq1.375\gev$
region, ACGL and CGL use it from threshold up to 2~\gev.

In fact, (\ref{eq:(1.1)}) gives a negative scattering length, whereas it is 
known that $a_2^{(2)}$ is {\sl positive},
 and it does not fit well the data below 0.7~\gev, as 
shown in \fig~2. 
Above 1~\gev, the modulus of (\ref{eq:(1.1)}) grows like $s$, whereas Regge 
theory requires all waves to tend to a multiple of $\pi$; i.e,
D2 should go to zero. 
In addition, we see in \fig~2 that  
Eq.(\ref{eq:(1.1)}) does not fit well 
the data between 1.4 and 2~\gev.
Finally, above 1.5~\gev, 
where the $\pi^-\pi^-\to\rho^-\rho^-$ channel opens, 
the D2 wave should be highly inelastic: but 
ACGL take it elastic up to 2~\gev.
The fact
 that the Regge representation of ACGL, CGL,
fits their sum rules is more  {\sl negative} than
 positive support.
For further detail, we refer to \cite{9}, PY1 and PY-FDR 
where we show that crossing sum rules are perfectly 
satisfied when considering the correct Reggeistics {\sl and} the 
correct D waves.

%\begin{figure}[htb]
\vspace*{.3cm}
{\psfig{file=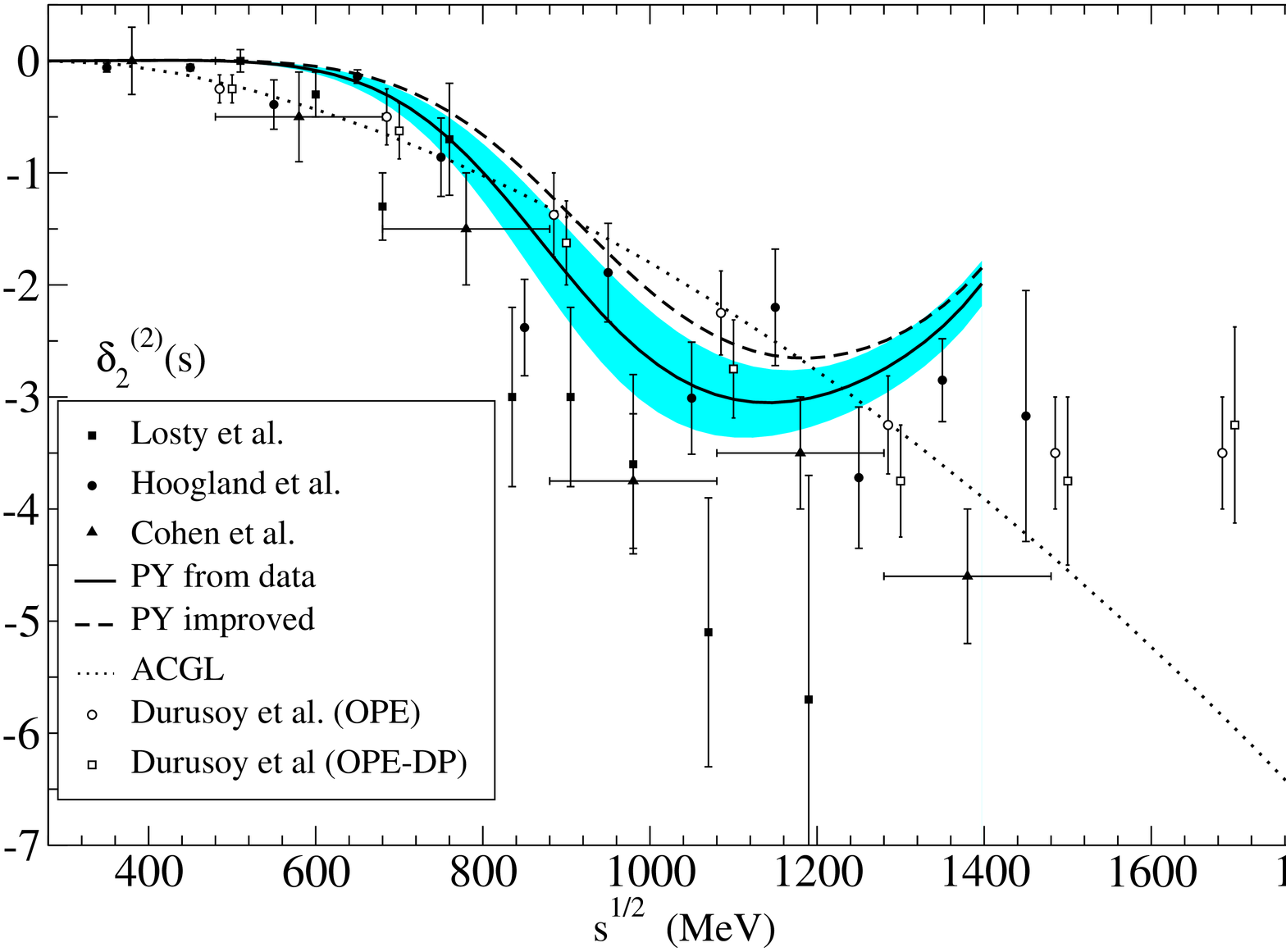,width=7.4cm,angle=-90}}
{\footnotesize {\bf Figure 2.}  $I=2$, $D$-wave phase
  shift and data \cite{13}. Continuous line: (PY-FDR) fit to data.
  Dashed line: (PY-FDR) improved parameters using dispersion
  relations. Dotted line: Martin, Morgan and Shaw fit which ACGL, CGL
  and CCGL use from threshold to $s^{1/2}=2\,\gev$.}
%\end{figure}
%\vspace*{.3cm}

\vspace*{-.2cm}
\section{CCGL claim the data are described by CGL but not by our 
 ``tentative solution"}

\noindent 
In PY1 we presented 
a ``tentative solution", merely a fit to low energy $\pi\pi$ data, 
in order to compare with the CGL low energy partial waves.
CCGL conclude that this tentative solution 
does not fit the experimental data, at least not as well as the phase shifts in CGL do. 
This is very surprising since,  our ``tentative solution" 
was obtained by just fitting data. 
We will discuss here the S0 wave, and particularly
the value of its phase shift at 800 MeV,  a key value 
in their input, and then the S2 wave.

\subsection{The S0 wave below 800~\mev}

\noindent
By looking at Fig.~1 in CCGL, reproduced here in Fig.~3a
only with the CGL and ``tentative solution'' of PY1,
it may seem that data fall 
on the CGL results, and are incompatible with the PY1
 ``tentative solution". 
However, {\it that only happens because CCGL have not plotted
all data}.
Fig.~1 of CCGL is certainly unfair with our tentative solution, 
which is a fit to an average of {\it published} data.
Indeed, in  \fig~3b  here, we include 
the solutions of different experimental analyses
\cite{15a,15b,15c,15d,15e,16} which 
ACGL, CGL quote in their references,
but CCGL do not show in their figure (we have also included 
in \fig~3b the recent data in \cite{15f}).
It is not clear why ACGL, CGL and CCGL
only consider a subset of all published data.

%\begin{figure}%[htb]
\vspace{.3cm}
\psfig{file=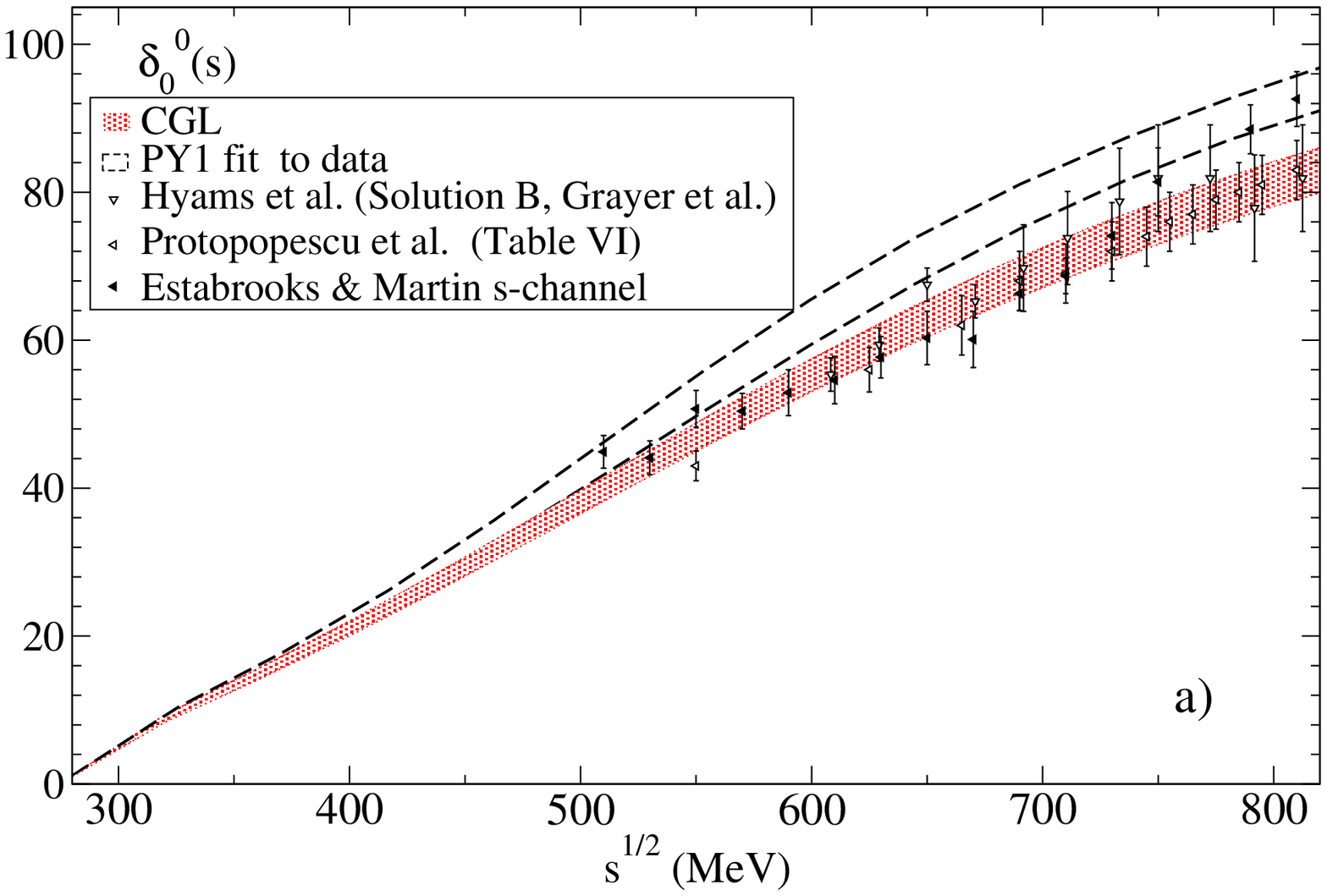,width=7.6cm,angle=-90}
\psfig{file=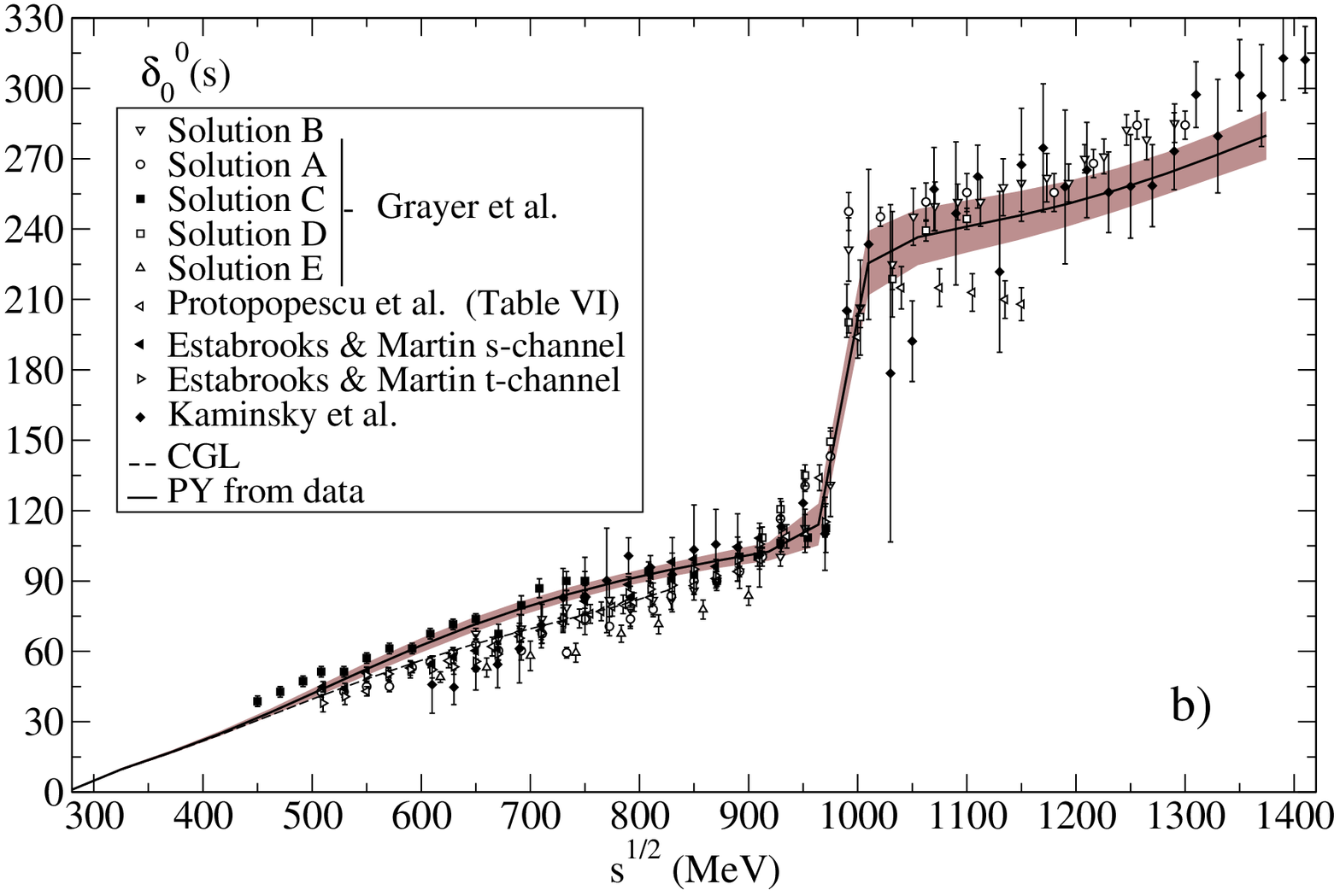,width=7.6cm,angle=-90}
{\footnotesize {\bf Figure 3.}  
S0 phase shifts. a) The shaded area is the CGL error 
band. The PY1 tentative 
solution lies between the dashed lines. Only  the
data included in 
\fig~1 of CCGL are shown.
b) PY1 Tentative solution and improved solution
of PY-FDR. The dashed line is the CGL phase. 
We show 
data from \cite{15a,15b,15c,15d,15e,15f,16}. (see PY-FDR for details)}
%\end{figure}
%\vspace{.1cm}

\subsection{The matching phase $\delta_0^{(0)}(0.8\,\gev)$}

\noindent
One of the most important input parameters \cite{10} in
ACGL, CGL is the S0
 phase shift at the matching point between low and high energy, $s_m\equiv
(0.8~\gev)^2$.  The data is affected by large systematic errors, but
ACGL consider the difference between S0 and P phase shifts
$\delta_1(s_m)-\delta_0^{(0)}(s_m)$
hoping some uncertainties may cancel.
In particular, by interpolation, they quote 
\begin{eqnarray}
&&24.8\pm3.8\degrees\quad\hbox{[Estabrooks and Martin, $s$-channel]}
%\nonumber 
\cr
&&30.3\pm3.4\degrees\quad\hbox{[Estabrooks and Martin, $t$-channel]}.
\cr
&&23.4\pm4.0\degrees\quad\hbox{[Hyams et al.]} 
\label{eq:(3.2)}
%\nonumber 
%}
\end{eqnarray}
and average them to obtain $26.6\degrees\pm3.7\degrees$. 

However, these three numbers stem 
from different analyses of the 
{\sl same experiment}, 
CERN-Munich \cite{15a,15e}, so that 
their spread measures the {\sl systematic} error.
Consequently, one should have enlarged the error
to cover {\sl all} central values:
\begin{eqnarray*}
\delta_1(s_m)-\delta_0^{(0)}(s_m)=
26.3\degrees\pm3.8\,({\rm sta.})\pm4.0\,({\rm sys.)}.
\end{eqnarray*}
Furthermore, since Estabrooks and Martin,\cite{15e},
 in their section 4, state that the input uncertainties, particularly
in the D-wave, lead to  {\it ``systematic changes in $\delta^0_S$
of the order of 10$\degrees$''}, that is, the $\pm4$ systematic error
is likely still optimistic.

Then ACGL average again their number with $26.5\pm4.2$, 
interpolated from Protopopescu
et al. \cite{16}
(Table VI), to obtain a surprisingly small
error:
$\delta_1(s_m)-\delta_0^{(0)}(s_m)\stackrel{ACGL}{=}26.6\degrees\pm2.8\degrees$.
However, in Protopopescu
et al. \cite{16}, we read that 
{\it ``the given errors...reflect only statistical error...and...should
be considered only as an indication of 
the minimum error in our computed values''}.
Still, CGL add back $\delta_1(s_m)=108.9\pm2\degrees$,  
and come up with,
\begin{equation}
\delta_0^{(0)}(0.8\,\gev)=82.3\pm3.4\degrees,
\label{eq:(3.1)}
\end{equation}
which is also used in CGL and defended in \cite{10}. 

To our view, the statistical and systematic errors should be added linearly,
in order to be conservative,
given the caveats of the original authors.
{\it But even this is most likely still optimistic},
since  there are more data, even more spreaded.
In fact, the value from Hyams et al., \cite{15a} quoted in 
(\ref{eq:(3.2)}), 
 is only one of {\sl five 
solutions published by the same experiment},
cf.~Grayer et al.\cite{15b}. The 
 datum of Protopopescu et al.,
is again one of the several analysis in \cite{16}, 
which differ among themselves 
again by  about $10\degrees$ --
not by chance in agreement with the statements quoted above
by  Estabrooks and Martin, \cite{15e}. 

%\begin{figure}%[htb]
\vspace{.3cm}
\psfig{file=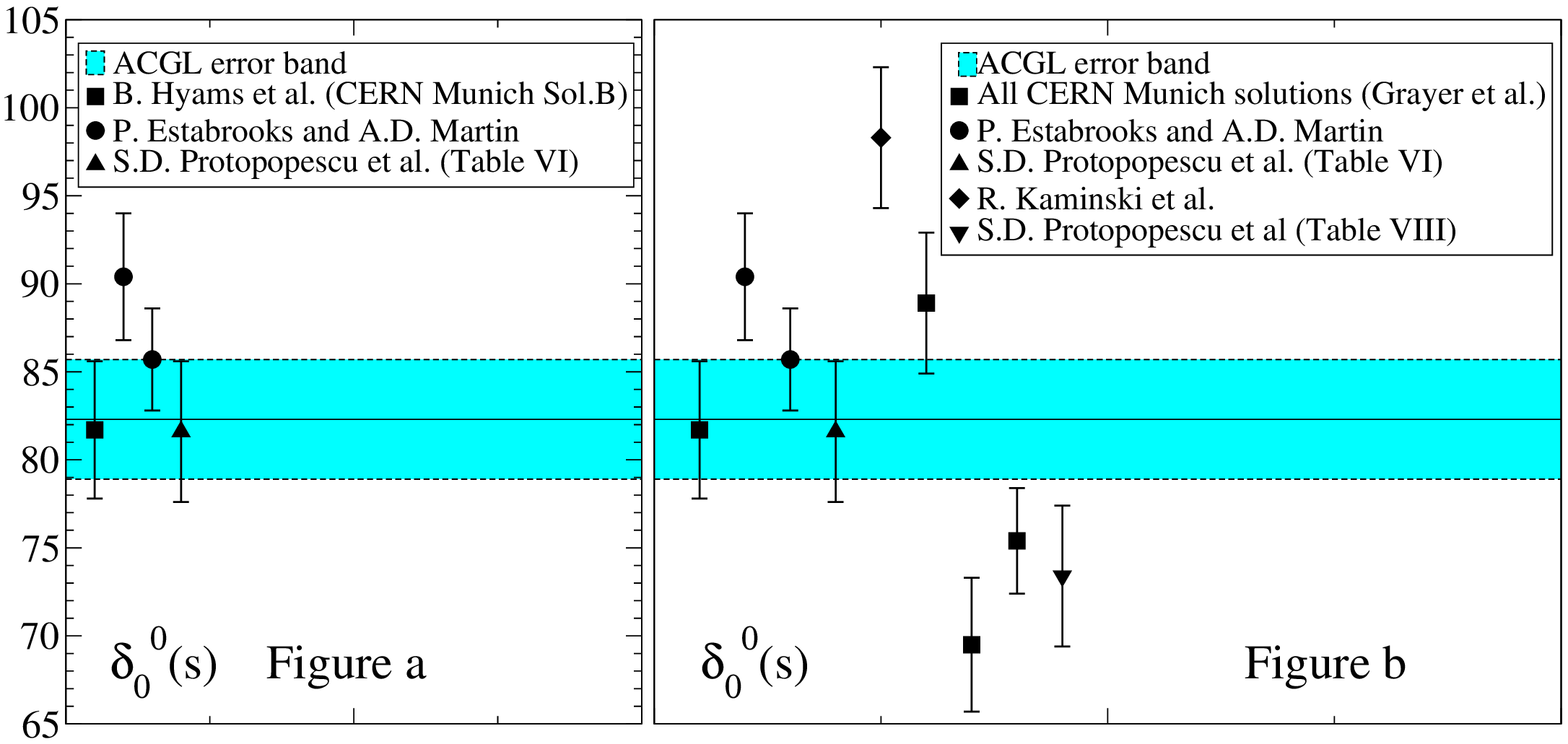,width=7.6cm,angle=-90}
{\footnotesize {\bf Figure 4.} a) Values of
$\delta_0^{(0)}(0.8\,\gev)$,
 as given by Colangelo in \fig~3 in ref~10. 
We include only data with error bars.
The shaded area covers the error band in ACGL, Eq.(\ref{eq:(3.1)}) here. Systematic errors
estimated \cite{15d} to be  $\sim10\degrees$ were not included.
b) Published data \cite{15a,15b,15c,15d,15e,15f,16}
on $\delta_0^{(0)}(0.8\,\gev)$. The band is like in 5.a. 
}
%\end{figure}
\vspace{.3cm}

%\begin{table}%[htbp]
%\vspace{.2cm}
\begin{center}
  \begin{tabular}{|c|c|}
\hline
Authors and references& $\delta_0^{(0)}(0.8\,\gev)$\\\hline
Kaminski et al. \cite{15f}&$98.3\pm5.3\degrees$\\
Grayer et al.,A \cite{15b}&$75.4\pm3.0\degrees$\\
Grayer et al.,B \cite{15b}&$81.7\pm3.9\degrees$\\
Grayer et al.,C \cite{15b}&$88.9\pm4.0\degrees$\\
Grayer et al.,E \cite{15b}&$69.5\pm3.8\degrees$\\
EM,s-channel \cite{15e}&$90.4\pm3.6\degrees$\\
EM,t-channel \cite{15e}&$85.7\pm2.9\degrees$\\
Protopopescu et al.\cite{16} {\footnotesize (VI)}&$81.6\pm4.0\degrees$\\
Protopopescu et al. \cite{16}{\footnotesize  (VIII)}&$73.4\pm4.0\degrees$\\
\hline
  \end{tabular}
\end{center}
\vspace{.1cm}
  {{\bf Table 1.} \footnotesize $\delta_0^{(0)}$ values 
{\it interpolated} to $0.8\,\gev$ from different experimental analysis.
ACGL, CGL and CCGL, take as
    experimental {\sl input },
    $\delta_0^{(0)}(0.8,\gev)=82.3\pm3.4\degrees$.
}
  \label{tab:800}
\vspace{.3cm}
%\end{table}

The ACGL, CGL matching input is summarized by Colangelo in \cite{10},
where in his Fig.~3 the 
ACGL uncertainty band is shown, but 
compared {\it only} with the data used in CGL and ACGL,
and {\it only with statistical errors}.
This can be seen in our Fig.4a above, where
we plot
the ACGL band, Eq.(\ref{eq:(3.1)}) here, with
the data included in \fig~3\ of \cite{10}.
Here we do not include the data that
does not have error bars, since, 
 they are not even used by ACGL to estimate their uncertainty.
In contrast, in \fig~4b we repeat \fig~4a, 
but we add other {\it published} data sets. 
We list the actual values and references in Table 1:
The small error band of ACGL does not
represent the experimental uncertainty.
Even less so when considering that many data come from different analyses
of the same experiment (like CERN-Munich and Estabrooks and Martin),
and their central value spread is not of statistical nature.

Finally, we want to remark that  without imposing
the ACGL, CGL surprisingly small uncertainty on 
the matching phase, a simultaneous fit \cite{Kaminski} to Roy equations and
 data yields  $\delta_0^{(0)}=92.6\degrees$
 whereas a fit only to data gives
$\delta_0^{(0)}=87.2\degrees$, %(they use $\delta_1=104.8\degrees$)
both
outside the value imposed by CGL and CGL as their input.

%\begin{figure}%[htb]
\vspace{.2cm}
\psfig{file=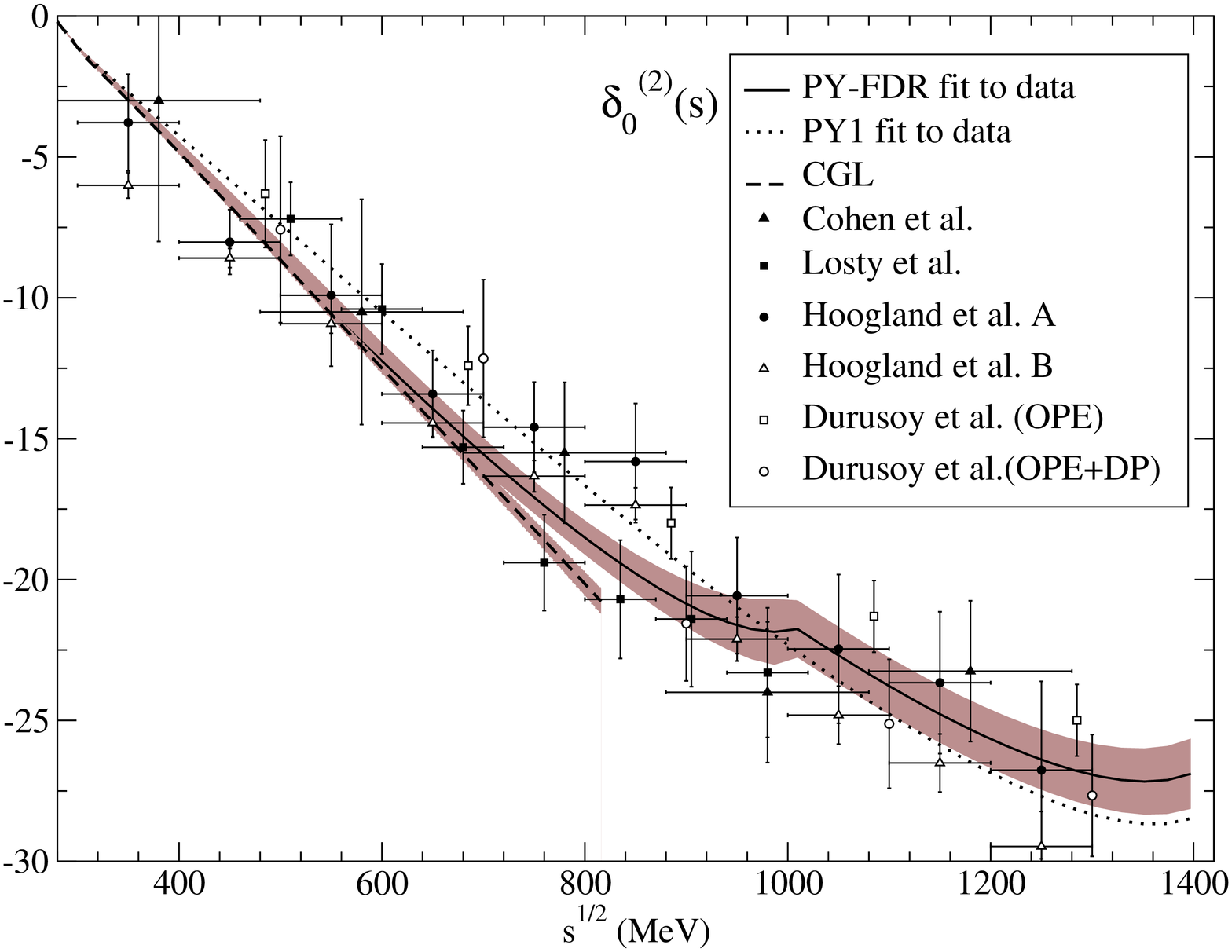,width=7.6cm,angle=-90}
{\footnotesize {\bf Figure 5.} The  
$I=2$, $S$-wave. Data comes from \cite{4,13}.
Continuous line: PY-FDR fit. 
Dotted line: PY1 fit. Data 
from Durusoy et al. and from Solution~B of 
Hoogland et al. were not included in the fits
(see PY-FDR for details).
The dashed line, below 
our fit, is the S2 phase of CGL \cite{2}.
The gray bands cover the respective uncertainties. }
%\end{figure}
%\vspace{.3cm}

\vspace{-.2cm}
\subsection{The S2 wave}

\noindent
The S2 phase shift data \cite{13} 
can be fitted with a simple
effective range expansion, like the
 ``tentative solution" of PY1, where we neglected 
the inelasticity below 1450~\mev, or, 
with greater precision, in PY-FDR, including the 
inelasticity above $\sim1\,\gev$. 
In \fig~5 the PY fit is shown to describe data better
than the CGL phase shift: 
not surprisingly since it was {\sl not} 
obtained from theory as the CGL solution. 
In fact, the  $\chi^2$ of the latter
gets  twice as large as that of the PY1 and PY-FDR
fits, as $\sqrt{s}$ tends to  800~\mev.
In addition, we show below that forward dispersion relations
are in conflict with a curve as low as that of CGL.

\section{
Regge formulas and threshold results}

\noindent
CCGL repeated the Roy equation analysis of CGL 
using the Regge PY1 formulas, and claimed that their results,
except for the S2 wave,
do not vary appreciably. Let us first note that
Roy equations use Regge expressions 
beyond their applicability region $|t|\ll s$ .
Second, that the effect of Regge formulas is strongly constrained
since CGL are forcing their
solutions to match the $\delta(s_m)$ phase shift
with an extremely small input uncertainty 
that, as we have seen in the previous section, 
basically neglects all systematic uncertainties.

However, low energy parameters can be calculated \cite{3}
with sum rules that involve only 
small values of $|t|$: First, $a(+0)\equiv a_2(\pi^0\pi^0\to\pi^+\pi^-)=2[a_2^{(0)}-a_2^{(2)}]/3$, 
and $a(00)\equiv a_2(\pi^0\pi^0\to\pi^0\pi^0)=2[a_2^{(0)}+2a_2^{(2)}]/3$
can be calculated with the Froissart Gribov (FG) representation, that needs only
information at $t=4M^2_\pi\simeq0.08\,\gev^2$.
Second, $2a_0^{(0)}-5a_0^{(2)}$, can be evaluated with the Olsson sum rule, 
that just needs $t=0$. We find that

i) Using inside the sum rule integrals the low energy 
parameterizations of 
CGL up to 800\mev, and experiment between 800 
and 1.42~\gev,  together with  the Regge parameters of PY1,
we  obtain $a(0+)=10.94\pm0.13$. Thus, the CGL result,
 $a(0+)=10.53\pm0.10$, obtained with a Wanders sum rule,
using their Regge and D-wave parametrizations instead, 
presents a $2.5\sigma$ mismatch. For the difference between the CGL 
calculation using Wanders sum rules minus the FG representation,
which cancels correlations, the mismatch is of
more than 4 standard deviations.
We agree with CCGL that this difference does not involve 
the S and P waves, and the mismatch is only due to the Regge and $l\geq2$ wave
input, different for CGL and PY. However, as we have just seen, it
 certainly affects
the $a(0+)$ total  value by about $2.5\sigma$.
The situation for $a(00)$ is very similar \cite{3}.

ii) 
From Eq.~(11.2) in CGL, we find,
in units of the pion mass,
\begin{equation}
2a_0^{(0)}-5a_0^{(2)}=0.663\pm0.006\quad \hbox{``CGL, direct''}
\label{eq:(2.2)}
\end{equation}
Alternatively, we can use the Olsson sum rule:
\begin{equation}
2a_0^{(0)}-5a_0^{(2)}=3M_\pi\int_{4M_\pi^2}^\infty \hbox{d} s\,
\frac{\imag F^{(I_t=1)}(s,0)}{s(s-4M_\pi^2)}.\label{eq:(2.3)}
\end{equation}
The total $I_t=1$ Regge contribution comes out the same
either with the PY1 parametrization, or with
the
improved PY-Regge and PY-FDR Regge parameters, i.e.
replacing
the contribution of the $\rho'$ trajectory and of the 
$\rho(1450)$ resonance for a slight increase in the 
rho residue to $\beta_\rho(0)=1.02\pm0.11$.
At low energy we use the 
S, P waves of CGL.  
We obtain $0.635\pm0.014$, in pion mass units.
There is a clear mismatch between the ``direct" 
result, Eq.(\ref{eq:(2.2)}), and the dispersive evaluations
of  $2a_0^{(0)}-5a_0^{(2)}$, which can be calculated 
with large precision. Concerning the S and P wave
scattering lengths individually, the PY and CGL results are in agreement,
mostly due to the larger PY error bars (
thus, our results do not question the large condensate scenario
of ChPT).

In summary, some quantities may not change 
if using different 
Regge asymptotics, but certainly others do.

\section{ CGL solutions and forward dispersion relations}

\noindent
In PY-FDR we studied a forward dispersion relation 
for the $I_t=1$, which  at threshold 
reduces to the Olsson
sum rule already evaluated in PY1. In addition,
we studied 
subtracted forward dispersion relations 
for $\pi^0\pi^+$ and $\pi^0\pi^0$ scattering.
Indeed, these relations imply the vanishing
of:
\begin{eqnarray}
&&\Delta_{1}\equiv\real
F^{(I_t=1)}(s,0)\label{(4.1a)}\nonumber\\
&&-\frac{2s-4M^2_\pi}{\pi}\pepe\int_{4M^2_\pi}^\infty\dd s'\,
\frac{\imag F^{(I_t=1)}(s',0)}{(s'-s)(s'+s-4M^2_\pi)},
\nonumber\\
&&\Delta_{00}\equiv\real F_{00}(s)-F_{00}(4M_{\pi}^2)-
\frac{s(s-4M_{\pi}^2)}{\pi}\times\label{(4.1b)}\nonumber\\
&&\pepe\int_{4M_{\pi}^2}^\infty\dd s'\,
\frac{(2s'-4M^2_\pi)\imag F_{00}(s')}{s'(s'-s)(s'-4M_{\pi}^2)(s'+s-4M_{\pi}^2)},
\nonumber\\
&&\Delta_{0+}\equiv\real F_{0+}(s)-F_{0+}(4M_{\pi}^2)-
\frac{s(s-4M^2_\pi)}{\pi}\times\label{(4.1c)}\nonumber\\
&&\pepe\int_{4M_{\pi}^2}^\infty\dd s'\,
\frac{(2s'-4M^2_\pi)\imag F_{0+}(s')}{s'(s'-s)(s'-4M_{\pi}^2)(s'+s-4M_{\pi}^2)}
\nonumber
\end{eqnarray}

We plot these differences in \fig~6, 
using  the CGL parameters for the S0, S2 and P waves below 800~\mev\,
where the mismatch is clearly seen. 
In contrast, as shown in 
PY-FDR, forward dispersion relations are 
fairly well satisfied below 800~\mev\ by simple fits to data on
 S0, S2 and P waves, and remarkably well
for our improved PY-FDR solutions.

%\begin{figure}%[htb]
  \vspace{.2cm} 
{\centering
\psfig{file=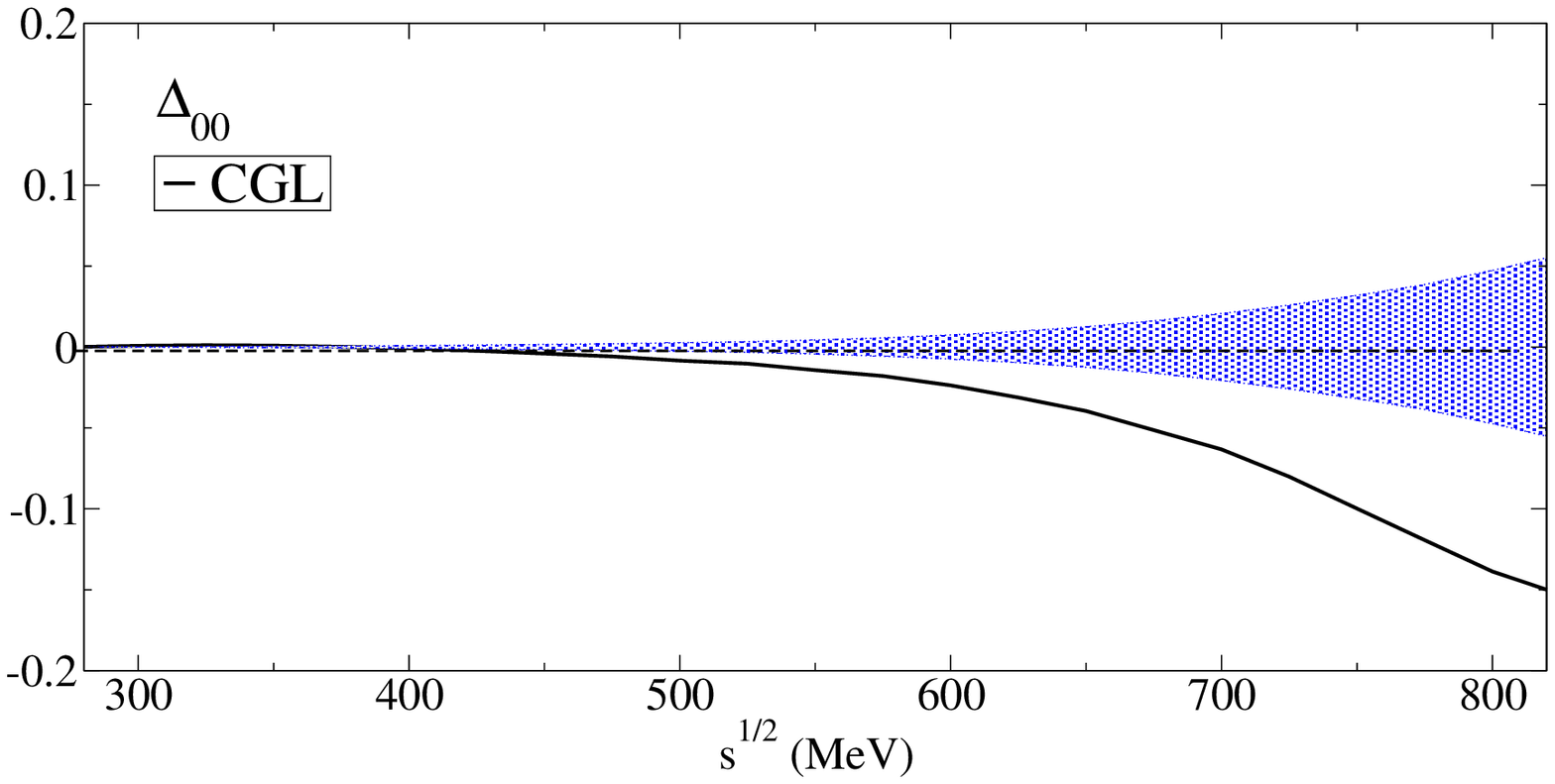,width=7.6cm,angle=-90}
  \vspace{-.32cm}
  \psfig{figure=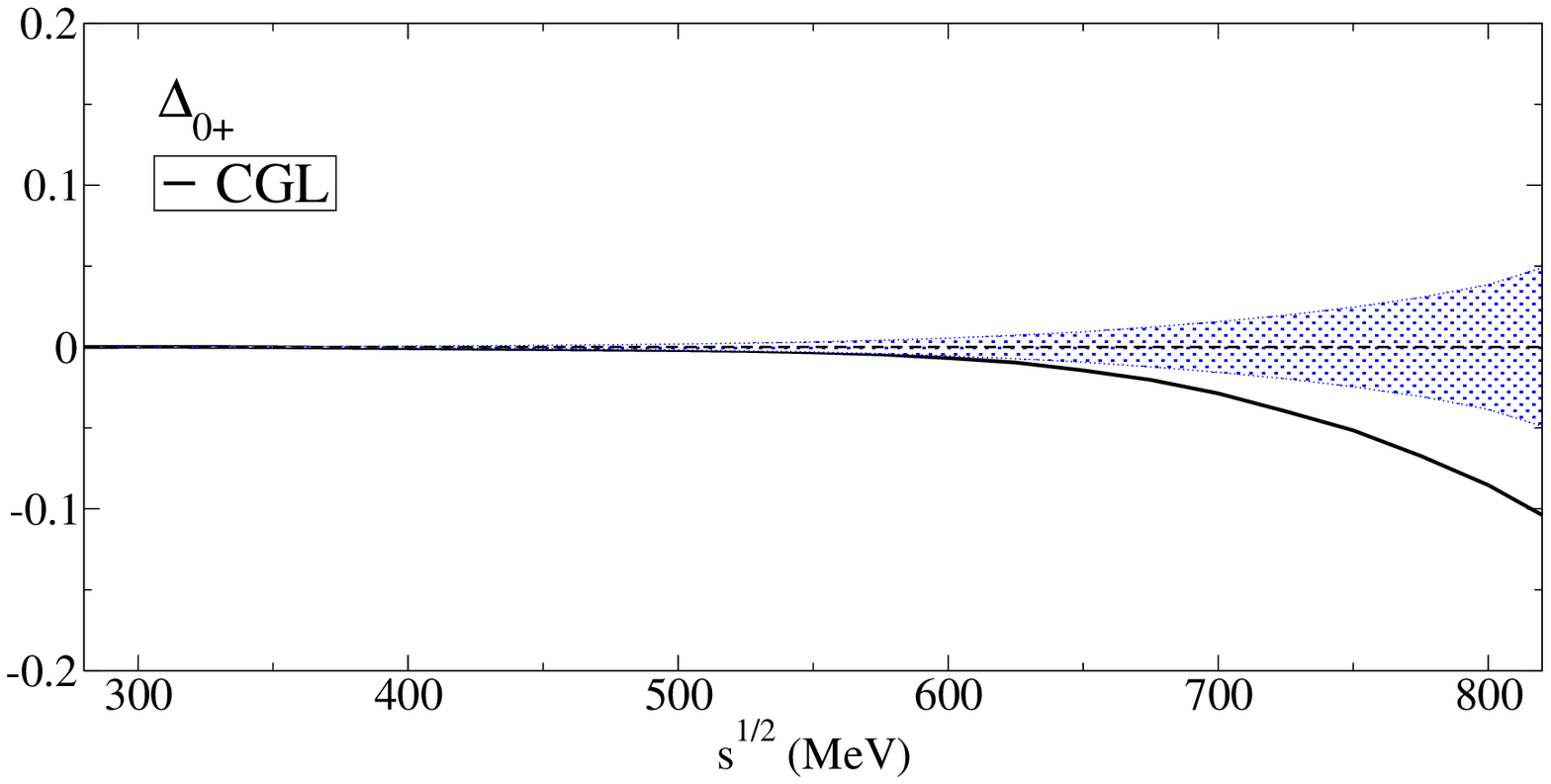,width=7.6truecm,angle=-90}
  \vspace{-.32cm}
  \psfig{figure=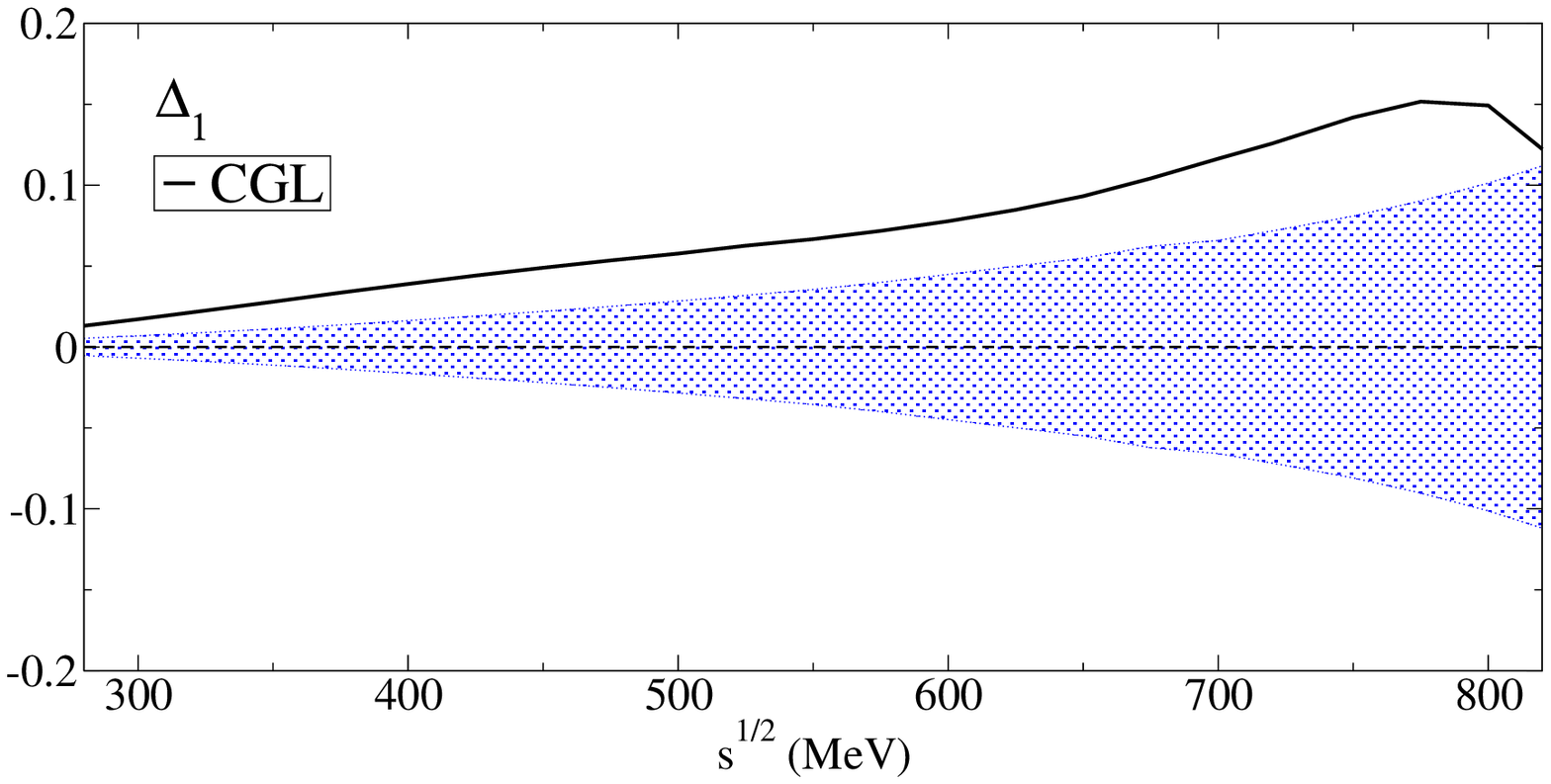,width=7.6truecm,angle=-90}
}
  {\footnotesize {\bf Figure 6} Continuos lines: for CGL S and P
phase shifts, the differences
    $\delta_i$ between the real parts calculated directly 
or from the dispersive formulas. Consistency occurs when
    curves fall within the shaded areas \cite{5}.}
%\end{figure}
\vspace{.1cm}

Finally, one may wonder why there is a mismatch for CGL,
since their S0 wave follows the Solution~B of Grayer et al.\cite{15b},
whose improved fit is fairly compatible with dispersion relations 
(with the correct Regge behaviour and 
D2 wave, see PY-FDR). 
However one should note that the Solution B 
fit, constrained to satisfy
dispersion relations, leads to an amplitude that 
violates the $\pi^0\pi^0$ dispersion relation 
at $s=2M^2_\pi$ (Table~3 in PY-FDR). 
In addition, it requires S2 phase shifts that 
disagree even more with that in CGL than the S2 wave obtained only 
by fitting data, as shown in \fig~7. Indeed, the S2 wave
obtained by CCGL from Roy equations but using PY asymptotics is also
displaced in the same direction.

%\begin{figure}%[htb]
\vspace{.2cm}
\psfig{file=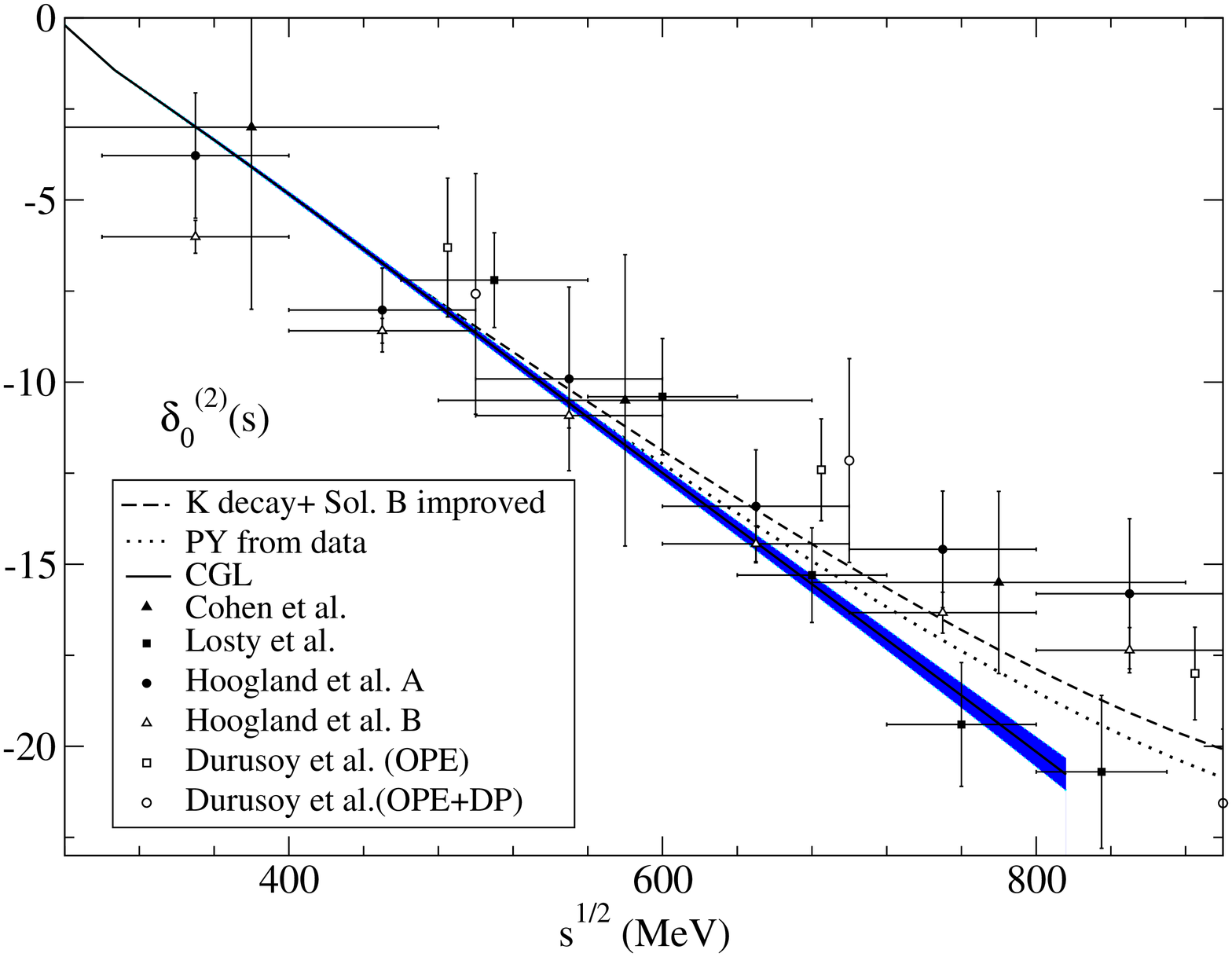,width=7.2cm,angle=-90}
{\footnotesize {\bf Figure 7} The  
$I=2$, $S$-wave phase shift. Data as in \fig~4.
Continuous line: CGL solution and error band. 
Dotted line: PY-FDR fit.
Dashed line: PY-FDR fit constrained
with dispersion relations when also improving the
S0 wave fit from Solution B of Grayer et al.\cite{15b}.}
%\end{figure}
%\vspace{.3cm}

\vspace{-.3cm}
\section{ The value of $b_1$.}

\noindent
In PY1 it was noted that the CGL 
value for the effective range parameter, $b_1=(5.67\pm0.13)\times10^{-3}\;M_{\pi}^{-5}$, was several standard 
deviations away from what one gets from the pion form factor \cite{7}
 $b_1=(4.73\pm0.23)\times10^{-3}\;M_{\pi}^{-5}$.
CCGL answered that such a result depended on the specific
form of the parametrization used for the P wave phase.
To clarify this matter, in PY-FDR we devised a fastly convergent sum rule 
that depends little on the high energy behaviour or the 
low-energy P wave phase shift, so it 
provides and independent determination of $b_1$:
\begin{eqnarray*}
M_\pi b_1&&=\,
\frac{2}{3}\int_{4M^2_\pi}^\infty\dd s\,\cr
&&\Bigg\{
\frac{1}{3}\left[\frac{1}{(s-4M^2_\pi)^3}-\frac{1}{s^3}\right]
\imag F^{(I_t=0)}(s,0)\cr
&&+\frac{1}{2}\left[\frac{1}{(s-4M^2_\pi)^3}+\frac{1}{s^3}\right]\imag F^{(I_t=1)}(s,0)\cr
&-&\,\frac{5}{6}\left[\frac{1}{(s-4M^2_\pi)^3}-
\frac{1}{s^3}\right]\imag F^{(I_t=2)}(s,0) \Bigg\}.
\end{eqnarray*}
The largest contribution comes
from S0 and P waves at low energy, while  
all other pieces (in particular, the Regge contributions) 
are substantially smaller than $10^{-3}$.  We find
\begin{equation}
b_1=(4.99\pm0.21)\times10^{-3}\;M_{\pi}^{-5}.
\label{eq:(5.1)}
\end{equation}
In PY-FDR, using fits constrained  with dispersion relations,
we found $b_1=(4.55\pm0.21)\times10^{-3}\;M_{\pi}^{-5}$.
In conclusion, all three experiment-based values for 
$b_1$ are fairly compatible among themselves, but
 several standard deviations
away from the CGL value.

Another matter addressed to by Colangelo in ref.~10 is that of the 
{\sl scalar} form factor of the 
pion. We will not discuss this here, but refer to 
the relevant literature \cite{17}.

\vspace{ -.2cm}
\section{Conclusions}
\vspace{ -.2cm}

In a series of works we have contested the input used by 
 Ananthanarayan, Colangelo, Gasser and Leutwyler \cite{1} (ACGL)
and Colangelo, Gasser and Leut\-wyler \cite{2} in their Roy equation
analysis of $\pi\pi$ scattering. This challenged the remarkable
precision claimed in CGL.
Subsequently, Caprini, Colangelo, Gasser and Leutwyler \cite{8,10} claimed 
to refute our arguments. Here, we collect the answers to their arguments,
clarifying the points of agreement and disagreement,
and showing {\sl experimental} 
support for our results. In particular:

i) We questioned the Regge formulas used in ACGL and CGL, 
which did not respect factorization.
Thus, we proposed a Regge parametrization, that, according
to 
Caprini, Colangelo, Gasser and Leutwyler \cite{8} (CCGL)
violated crossing  symmetry rather strongly.
Later on, we ``rediscovered'' the existing data on $\pi\pi$ total cross sections:
they turned out to be well described
%which are well described 
with our Regge formulas,
but not with those of ACGL and CGL. 
Furthermore, and although the issue is now irrelevant because there is data to compare with, 
we have shown that crossing is satisfied with our Regge equations.

We have also shown that the D2 wave parameterization in ACGL, CGL,
derived in the seventies from intermediate energy data,
was used outside its applicability range.

ii) Since the CGL phase shifts did not satisfy a number of sum rules
 when using a D2 wave and a Regge description 
compatible with data, we proposed a ``tentative solution'' \cite{3},
from a fit to an average of data. Surprisingly, CCGL 
claimed that this ``tentative solution'' did not describe experiment. 
We have shown here that this only happens
because ACGL, CGL and CCGL only consider a subset of all published data.

iii) We have shown here that
the uncertainty imposed by ACGL and CGL on the
difference between P and S0 phase shifts
at 800 MeV, is largely underestimated when taking into account
 all published data 
and their systematic uncertainties, that are emphasized
in some of the original references.
This is an important input for their Roy equation analysis and,
as remarked by CCGL,  the main 
source
of error for the Olsson sum rule.

iv) CCGL claim that the CGL Roy equation analysis remains unchanged
(except for the S2 wave) if using our Regge asymptotics. We agree that this 
can be the case for certain observables, like the S and P wave scattering lengths
(which are indeed compatible with our values).
However, other quantities, like the phase shifts 
at intermediate energy or
the D-wave scattering lengths can vary considerably.

v) We have confirmed our result for the $b_1$ P-wave threshold parameter
with a new sum rule that depends little on Regge asymptotics, or the precise form
of the P wave parametrization.
 
vi) Finally, we have recently shown that 
the CGL S and P phase shifts fail to satisfy three forward dispersion relations 
up to $\sqrt{s}\leq800\, \mev$ by several standard deviations.

Apart from this discussion, 
we also used  forward dispersion
relations to check the consistency \cite{5} of different data sets,
including the precise data from kaon decays. As it
is well known, there are many different phase shift analyses, often
incompatible even within the same experiment. Surprisingly, we 
found that some of the most
frequently used data sets are inconsistent 
with forward dispersion relations and sum rules, 
and should therefore be used with caution.
Finally, these forward dispersion relations and sum rules 
were used to constrain the
data fits and to provide a consistent amplitude
 easy to implement by those interested
in $\pi\pi$ scattering.

\vspace{ -.2cm}
\section*{Acknowledgments}
\vspace{ -.2cm}
\noindent
Work partially supported by DGICYT Spain, contracts
FPA2005-02327 and BFM2002-01868, and the EURIDICE network 
contract HPRN-CT-2002-00311 as well as the EU Hadron Physics Project, 
contract number RII3-CT-2004-506078.

\vspace{ -.2cm}
\bibliography{apssamp}% Produces the bibliography via BibTeX.
%\vspace{ .6cm}

\end{document}